\documentclass[letterpaper]{ptephy}

\usepackage{bm}
\usepackage{color}
\usepackage{ulem}

\renewcommand{\sout}{\bgroup \color[rgb]{1,0,0} \ULdepth=-.5ex \ULset}

\usepackage{times}
\usepackage{multirow}

\newcommand{\Psfig}[2]{\includegraphics[width=#1]{#2}}

\def\mev{\text{ MeV}}
\def\gev{\text{ GeV}}

\def\naive{na\"{i}ve }

\begin{document}

\title{$\Xi (1690)$ as a $\bar{K} \Sigma$ molecular state}

\author{Takayasu Sekihara}

\address{Research Center for Nuclear Physics (RCNP), Osaka University,
  Ibaraki, Osaka, 567-0047, Japan}

\begin{abstract}%
  We show that a $\Xi ^{\ast}$ pole can be dynamically generated near
  the $\bar{K} \Sigma$ threshold as an $s$-wave $\bar{K} \Sigma$
  molecular state in a coupled-channels unitary approach with the
  leading-order chiral interaction.  This $\Xi ^{\ast}$ state can be
  identified with the $\Xi (1690)$ resonance with $J^{P} = 1/2^{-}$.
  We find that the experimental $\bar{K}^{0} \Lambda$ and $K^{-}
  \Sigma ^{+}$ mass spectra are qualitatively reproduced with the $\Xi
  ^{\ast}$ state.  Moreover we theoretically investigate properties of
  the dynamically generated $\Xi ^{\ast}$ state.
\end{abstract}


\parindent0pt

\maketitle

\section{Introduction}

Investigating the internal structure of hadrons is one of the most
important subjects in hadron physics.  The motivation for the
investigation is that we expect the existence of exotic hadrons, which
are not able to be classified as $q q q$ for baryons nor $q \bar{q}$
for mesons.  Actually, the fundamental theory of strong interaction,
QCD, does not prohibit such exotic systems as long as they are color
singlet, and there are indeed several exotic hadron candidates which
cannot fit into the classifications by the constituent quark
models~\cite{Agashe:2014kda}.  In order to clarify the internal
structure of exotic hadron candidates and to discover genuine exotic
hadrons, great efforts have been continuously made in both
experimental and theoretical sides.  In this context, it is very
encouraging that charged quarkonium-like states and
charmonium-pentaquark states were observed in the heavy quark sector
by Belle~\cite{Belle:2011aa} and by LHCb~\cite{Aaij:2015tga},
respectively.

In this Article we focus on the $\Xi (1690)$ resonance and
theoretically investigate its structure in terms of the $\bar{K}
\Sigma$ component.  Historically this resonance was discovered as a
threshold enhancement in both the neutral and charged $\bar{K} \Sigma$
mass spectra in the $K^{-} p \to (\bar{K} \Sigma) K \pi$ reaction at
$4.2 \gev /c$~\cite{Dionisi:1978tg}.  Several experimental and
theoretical studies have followed in, {\it e.g.},
Refs.~\cite{Biagi:1981cu, Biagi:1986zj, Adamovich:1997ud, Abe:2001mb,
  Link:2005ut, Aubert:2008ty, Ablikim:2015apm} and
Refs.~\cite{Chao:1980em, Capstick:1986bm, Glozman:1995fu,
  GarciaRecio:2003ks, Oh:2007cr, Pervin:2007wa, Gamermann:2011mq,
  Sharma:2013rka, Xiao:2013xi}, respectively, and today the $\Xi
(1690)$ resonance is attributed a three-star status in the Particle
Data Group table~\cite{Agashe:2014kda}.  Its mass and width are $1690
\pm 10 \mev$ and $< 30 \mev$, respectively~\cite{Agashe:2014kda}, but
a relatively narrow width, {\it e.g.}, $10 \pm 6
\mev$~\cite{Adamovich:1997ud} was reported as well.  In addition, the
small ratio of the $\Xi (1690)$ branching fractions $\Gamma (\pi \Xi)
/ \Gamma (\bar{K} \Sigma) < 0.09$ has been
observed~\cite{Agashe:2014kda}.  Its spin/parity has been expected to
be $J^{P} = 1/2^{-}$ from the beginning~\cite{Dionisi:1978tg}, and
this was supported by a recent experiment~\cite{Aubert:2008ty}.  Then,
a difficulty emerges; assuming $J^{P} = 1/2^{-}$, $\Xi (1690)$ couples
to the $\pi \Xi$ channel in $s$ wave, and hence $\Xi (1690)$, as a $q
q q$ state, should inevitably decay to the $\pi \Xi$ channel to some
extent in a \naive quark model, which contradicts the above
experimental results and implications.  This implies that $\Xi (1690)$
might have some nontrivial structure than usual $q q q$ state.

In this study, in order to describe $\Xi (1690)$, we perform a
coupled-channel analysis of the $s$-wave $\bar{K} \Sigma$, $\bar{K}
\Lambda$, $\pi \Xi$, and $\eta \Xi$ scatterings.  For this purpose we
employ the so-called chiral unitary approach~\cite{Kaiser:1995eg,
  Oset:1997it, Oller:2000fj, Lutz:2001yb, Jido:2003cb}, which is
formulated with the meson--baryon coupled-channels scattering equation
in an algebraic form based on the combination of the chiral
perturbation theory and the unitarization of the scattering amplitude.
One of the most remarkable properties of this approach is that a
simple driving term, or interaction kernel, provided by the chiral
Lagrangian with a small number of free parameters can reproduce
experimental observables such as cross sections fairly well.  The most
important application of the chiral unitary approach is the
description of the $\Lambda (1405)$ resonance~\cite{Hyodo:2011ur}.  In
this Article we apply the chiral unitary approach to the strangeness
$S = -2$ sector, where $\Xi ^{\ast}$ resonances exist.  In the chiral
unitary approach the $S = -2$ sector was already studied in
Refs.~\cite{Ramos:2002xh, GarciaRecio:2003ks, Gamermann:2011mq}.  On
the one hand, in Ref.~\cite{Ramos:2002xh} a $\Xi ^{\ast}$ state below
the $\bar{K} \Lambda$ threshold was discussed and identified with $\Xi
(1620)$.  On the other hand, in Refs.~\cite{GarciaRecio:2003ks,
  Gamermann:2011mq} the authors obtained several $\Xi ^{\ast}$ poles
such as $\Xi (1620)$, $\Xi (1690)$, and $\Xi (1950)$ together with
many hadronic resonances in $S = 0$ to $-3$.  Especially in
Ref.~\cite{GarciaRecio:2003ks} they found that in a flavor $SU(3)$
symmetric world $\Xi (1690)$ is a member of two degenerated octets,
which also contain one of the two-$\Lambda (1405)$ poles coming from
the $\bar{K} N$ bound state, $N (1535)$, and so on.  In this study we
extend the analyses in Refs.~\cite{Ramos:2002xh, GarciaRecio:2003ks}
by concentrating on the phenomena near the $\bar{K} \Sigma$ threshold
and on $\Xi (1690)$.  In the following we will show that the chiral
unitary approach in $S = -2$ can qualitatively reproduce the
experimental data of the $\bar{K}^{0} \Lambda$ and $K^{-} \Sigma ^{+}$
mass spectra, dynamically generate a $\Xi (1690)$ pole as an $s$-wave
$\bar{K} \Sigma$ molecular state near the $\bar{K} \Sigma$ threshold,
and naturally explain the decay properties of $\Xi (1690)$.

\section{Formulation}

First of all we formulate the meson--baryon scattering amplitude $T_{j
  k} ( w )$ in $s$ wave in the chiral unitary approach, where $w$ is
the center-of-mass energy and $j$ and $k$ are the channel indices.
The scattering amplitude is the solution of the Bethe--Salpeter
equation in a coupled-channels algebraic form
\begin{equation}
T_{j k} ( w ) 
= V_{j k} ( w )
+ \sum _{l} V_{j l} ( w ) G_{l} ( w ) T_{l k} ( w ) ,
\label{eq:BSeq}
\end{equation}
with the interaction kernel $V_{j k}$ taken from the chiral
perturbation theory and the meson--baryon loop function $G_{j}$.  The
treatment of the algebraic form was justified first by the so-called
on-shell factorization~\cite{Oset:1997it} and then by the dispersion
relation and the $N/D$ method~\cite{Oller:2000fj}.  In this study we
consider the system with $S = -2$ and charge $Q = 0$ or $-1$, where we
take into account six two-body channels ($K^{-} \Sigma ^{+}$,
$\bar{K}^{0} \Sigma ^{0}$, $\bar{K}^{0} \Lambda$, $\pi ^{+} \Xi ^{-}$,
$\pi ^{0} \Xi ^{0}$, and $\eta \Xi ^{0}$) for the neutral states and
similarly six channels ($\bar{K}^{0} \Sigma ^{-}$, $K^{-} \Sigma
^{0}$, $K^{-} \Lambda$, $\pi ^{-} \Xi ^{0}$, $\pi ^{0} \Xi ^{-}$, and
$\eta \Xi ^{-}$) for the charged states.  These channels are labeled
by the indices $j = 1$, ... , $6$ in the above orders for the neutral
and charged states, respectively.  For the interaction kernel $V_{j
  k}$ we use the leading-order chiral perturbation theory in $s$ wave,
i.e., the Weinberg--Tomozawa interaction.  After the projection to the
$s$ wave, the interaction is expressed as
\begin{equation}
V_{j k} ( w ) = - \frac{C_{j k}}{4 f_{j} f_{k}} 
( 2 w - M_{j} - M_{k} ) 
\sqrt{\frac{E_{j} + M_{j}}{2 M_{j}}}
\sqrt{\frac{E_{k} + M_{k}}{2 M_{k}}} , 
\label{eq:V_WT}
\end{equation}
with the $j$th channel meson decay constant $f_{j}$, baryon energy
$E_{j} \equiv ( s + M_{j}^{2} - m_{j}^{2} ) / (2 w)$, the squared
center-of-mass energy $s \equiv w^{2}$, and the baryon and meson
masses $M_{j}$ and $m_{j}$ in the $j$th channel, respectively.  In
this study we use the physical masses unless explicitly mentioned.
The factor $C_{j k}$ is the Clebsch--Gordan coefficient determined
from the group structure of the flavor $SU(3)$ symmetry and its value
is listed in Table~\ref{tab:coff0}.  The meson decay constants are
chosen at their physical values~\cite{Agashe:2014kda}: $f_{\pi} = 92.2
\mev$, $f_{K} = 1.2 f_{\pi}$, and $f_{\eta} = 1.3 f_{\pi}$.  Since
$V_{j k}$ in Eq.~\eqref{eq:V_WT} depends only on the center-of-mass
energy $w$, we can put $V_{j k}$ out of the loop integral in the
scattering equation, which hence becomes an algebraic form in
Eq.~\eqref{eq:BSeq}.  For the meson--baryon loop function $G_{j} (w)$,
we take a covariant expression and calculate the integral with the
dimensional regularization.  As a result, the loop function depends on
a subtraction constant $a_{j} (\mu _{\rm reg})$ in each channel at the
regularization scale $\mu _{\rm reg}$.  The explicit expression of the
loop function can be found in Ref.~\cite{Sekihara:2010uz}.  An
important point to be noted is that the interaction kernel does not
contain free parameters at the present order and its strength is fixed
entirely by the coefficients $C_{j k}$ and the meson decay constants.
Therefore, only the subtraction constant in each channel is the model
parameter.  In this study we assume the isospin symmetry for the
subtraction constants, {\it e.g.}, $a_{\bar{K} \Sigma} = a_{K^{-}
  \Sigma ^{+}} = a_{\bar{K}^{0} \Sigma ^{0}}$, so we will have four
model parameters in neutral and charged $S = -2$ systems,
respectively.

\def\arraystretch{1.1}

\begin{table}[!t]
  \caption{Coefficients $C_{j k} = C_{k j}$ for the channels in $Q =
    0$ and $S = -2$.  From these values we can obtain the coefficients
    for the channels in $Q = -1$ and $S = -2$ by using the relation
    $C_{j k} (Q = -1, \, S = -2) = \xi _{j} \xi _{k} C_{j k} (Q = 0, \,
    S = -2)$ with $\xi _{1} = \xi _{3} = \xi _{4} = \xi _{6} = +1$
    and $\xi _{2} = \xi _{5} = -1$. }
  \label{tab:coff0}
  \centering
  \begin{tabular}{l|cccccc}
    \hline
    \hline
    & $K^{-} \Sigma ^{+}$ & $\bar{K}^{0} \Sigma ^{0}$ & $\bar{K}^{0} \Lambda$ 
    & $\pi ^{+} \Xi ^{-}$ & $\pi ^{0} \Xi ^{0}$ & $\eta \Xi ^{0}$ \\
    \hline
    $K^{-} \Sigma ^{+}$ & 
    $1$ & $- \sqrt{2}$ & $0$ & $0$ & $- 1/\sqrt{2}$ & $- \sqrt{3/2}$ 
    \\
    $\bar{K}^{0} \Sigma ^{0}$ & 
    $- \sqrt{2}$ & $0$ & $0$ & $- 1/\sqrt{2}$ & $-1/2$ & $\sqrt{3/4}$ 
    \\
    $\bar{K}^{0} \Lambda$ & 
    $0$ & $0$ & $0$ & $- \sqrt{3/2}$ & $\sqrt{3/4}$ & $-3/2$ 
    \\ 
    $\pi ^{+} \Xi ^{-}$ & 
    $0$ & $- 1/\sqrt{2}$ & $- \sqrt{3/2}$ & $1$ & $- \sqrt{2}$ & $0$
    \\
    $\pi ^{0} \Xi ^{0}$ & 
    $- 1/\sqrt{2}$ & $-1/2$ & $\sqrt{3/4}$ & $- \sqrt{2}$ & $0$ & $0$
    \\
    $\eta \Xi ^{0}$ & 
    $- \sqrt{3/2}$ & $\sqrt{3/4}$ & $-3/2$ & $0$ & $0$ & $0$ \\
    \hline
    \hline
  \end{tabular}
\end{table}

When the interaction is enough attractive, the interaction can
dynamically generate a pole of the scattering amplitude for a
resonance or a bound state.  The pole is characterized by the pole
position $w_{\rm pole}$ and its residue $g_{j} g_{k}$:
\begin{equation}
T_{j k} ( w ) = \frac{g_{j} g_{k}}{w - w_{\rm pole}}
+ ( \text{regular at }w = w_{\rm pole} ) . 
\label{eq:amp_pole}
\end{equation}
The constant $g_{j}$ can be interpreted as the coupling constant of
the resonance to the $j$th two-body channel.  The pole position and
residue reflect the structure of the resonance.  Recently this
statement is formulated in terms of the
compositeness~\cite{Hyodo:2011qc, Hyodo:2013nka}.  First it was shown
in Refs.~\cite{Gamermann:2009uq, YamagataSekihara:2010pj} that the
$j$th channel two-body wave function is proportional to the coupling
constant $g_{j}$ for an energy independent separable interaction, and
then the case of a general separable interaction, including the
present formulation, was studied in Ref.~\cite{Sekihara:2014kya}.  In
the present formulation, we can calculate the $j$th channel two-body
wave function for the resonance as~\cite{Sekihara:2014kya}
\begin{equation}
  \tilde{\Psi}_{j} ( \bm{q} ) = 
  \frac{g_{j} \sqrt{4 M_{j} w_{\rm pole}}}
  {w_{\rm pole}^{2} - [ \omega _{j} (\bm{q}) + \Omega _{j} (\bm{q})]^{2}} ,
  \quad 
  \omega _{j} (\bm{q}) \equiv \sqrt{m_{j}^{2} + \bm{q}^{2}} , 
  \quad 
  \Omega _{j} (\bm{q}) \equiv \sqrt{M_{j}^{2} + \bm{q}^{2}} , 
\end{equation}
with the relative momentum of the state $\bm{q}$, and the $j$th
channel compositeness $X_{j}$ is obtained as the norm of the $j$th
channel two-body wave function as~\cite{Sekihara:2014kya}
\begin{equation}
  X_{j} \equiv \int \mathcal{D} \bm{q} 
  \left [ \tilde{\Psi}_{j} ( \bm{q} ) \right ]^{2}  
  = - g_{j}^{2} \left [ \frac{d G_{j}}{d w} \right ] _{w = w_{\rm pole}} , 
  \quad 
  \mathcal{D} \bm{q} \equiv \frac{d^{3} q}{( 2 \pi )^{3}} 
  \frac{\omega _{j} (\bm{q}) + \Omega _{j} (\bm{q})}
  {2 \omega _{j} (\bm{q}) \Omega _{j} (\bm{q})} .
\end{equation}
where the measure $\mathcal{D} \bm{q}$ guarantees the Lorentz
invariance of the integral and we have transformed the integral into
the derivative of the loop function (for details of the calculation,
see Ref.~\cite{Sekihara:2014kya}).  Here we note that we do not
calculate the absolute value squared but the complex number squared of
$\tilde{\Psi}_{j} (\bm{q})$ since we employ the Gamow vector for the
resonance so as to obtain the correct normalization of the resonance
wave function~\cite{Sekihara:2014kya}.  In addition to the
compositeness, one can calculate the elementariness $Z$ as the
contributions from implicit channels, which do not appear as explicit
degrees of freedom in the practical model space, such as compact $q q
q$ states.  Namely, on the assumption that the energy dependence of
the interaction originates from implicit channels, the elementariness
is expressed as~\cite{Sekihara:2014kya}
\begin{equation}
Z = - \sum _{j, k} g_{k} g_{j} \left [ G_{j} \frac{d V_{j k}}{d w} 
G_{k} \right ] _{w = w_{\rm pole}} .
\end{equation}
Then it is important that the sum of the compositeness and
elementariness coincides with the normalization of the total wave
function for the resonance $| \Psi \rangle$ and is exactly
unity~\cite{Sekihara:2014kya}:
\begin{equation}
\langle \Psi ^{\ast} | \Psi \rangle = \sum _{j} X_{j} + Z = 1 ,
\label{eq:norm}
\end{equation}
where the bra state $\langle \Psi ^{\ast} |$ has been used to
correctly normalize the resonance wave function in terms of the Gamow
vector.  The condition of the correct normalization as
unity~\eqref{eq:norm} is guaranteed by a generalized Ward identity
proved in Ref.~\cite{Sekihara:2010uz}.  We note that in general both
the compositeness $X_{j}$ and elementariness $Z$ are not observable
and hence are model dependent quantities.  Furthermore, they become
complex for a resonance state, so we cannot interpret the
compositeness (elementariness) as the probability of finding a
two-body (implicit) component inside the resonance.  However, based on
the normalization~\eqref{eq:norm}, we can interpret it for a resonance
with a wave function similar to a bound state one, as for the $\Xi
^{\ast}$ resonance in the following discussions.

\section{Numerical results}

Now we solve the scattering equation to obtain the scattering
amplitude $T_{j k} (w)$ and show the numerical results.  In the
following we mainly consider the neutral charge system since the
experimental data on both the $\bar{K}^{0} \Lambda$ and $K^{-} \Sigma
^{+}$ mass spectra are available~\cite{Abe:2001mb}.

\def\arraystretch{1.0}

\begin{table}[!t]
  \caption{Parameter sets~$\Lambda$, $\Sigma$, $\Xi$, and Fit, and
    properties of the neutral $\Xi (1690)$ state.  The regularization
    scale is $\mu _{\rm reg} = 630 \mev$ in all channels.  We also
    show the $\chi ^{2}$ value for the $\bar{K}^{0} \Lambda$ and
    $K^{-} \Sigma ^{+}$ mass spectra divided by the number of degrees
    of freedom, $\chi ^{2} / N_{\rm d.o.f.}$, and the ratio of the two
    branching fractions $R$ defined in Eq.~\eqref{eq:R}.}
  \label{tab:neutral}
  \centering
  \begin{tabular}{lcccc}
    \hline
    \hline
    & Set~$\Lambda$ & Set~$\Sigma$ & Set~$\Xi$ & Fit \\
    \hline
    $a_{\bar{K} \Sigma}$ 
    & $-2.30$
    & $-2.23$
    & $-2.10$
    & $-1.98$ 
    \\
    $a_{\bar{K} \Lambda}$ 
    & $-2.15$   
    & $-2.07$
    & $-1.91$
    & $-2.07$
    \\
    $a_{\pi \Xi}$ 
    & $-2.08$
    & $-1.99$
    & $-1.77$
    & $-0.75$
    \\
    $a_{\eta \Xi}$ 
    & $-2.57$
    & $-2.52$
    & $-2.43$
    & $-3.31$\vspace{2pt} 
    \\
    $\chi ^{2} / N_{\rm d.o.f.}$ 
    & $65.3 / 57$
    & $66.6 / 57$
    & $81.2 / 57$
    & $59.0 / 57$\vspace{2pt} 
    \\
    $R$ & $1.32$ & $3.04$ & $6.07$ & $1.06$
    \\
    \hline
    $w_{\rm pole}$ & $1682.6 - 0.8 i \mev$
    & \multirow{14}{*}{\rotatebox{-30}{No $\Xi (1690)$ pole}}
    & \multirow{14}{*}{\rotatebox{-30}{No $\Xi (1690)$ pole}}
    & $1684.3 - 0.5 i \mev$\vspace{3pt} \\
    $g_{K^{-} \Sigma ^{+}}$ & $1.00 + 0.22 i$
    & &
    & $1.02 + 0.60 i$
    \\
    $g_{\bar{K}^{0} \Sigma ^{0}}$ & $-0.73 - 0.15 i \phantom{-}$ 
    & &
    & $-0.76 - 0.41  i \phantom{-}$
    \\
    $g_{\bar{K}^{0} \Lambda}$ & $0.24 - 0.04 i$ 
    & &
    & $0.38 + 0.20 i$
    \\
    $g_{\pi ^{+} \Xi ^{-}}$ & $0.04 + 0.08 i$ 
    & &
    & $0.06 - 0.05 i$
    \\
    $g_{\pi ^{0} \Xi ^{0}}$ & $-0.05 - 0.05 i \phantom{-}$ 
    & &
    & $-0.09 + 0.05 i \phantom{-}$
    \\
    $g_{\eta \Xi ^{0}}$ 
    & $-0.76 - 0.17 i \phantom{-}$ 
    & &
    & $-0.66 - 0.48 i \phantom{-}$\vspace{4pt} \\
    $X_{K^{-} \Sigma ^{+}}$ 
    & $0.77 - 0.10 i$ 
    & &
    & $0.83 - 0.31 i$
    \\
    $X_{\bar{K}^{0} \Sigma ^{0}}$ & $0.12 + 0.04 i$ 
    & &
    & $0.12 + 0.17 i$
    \\
    $X_{\bar{K}^{0} \Lambda}$ & $0.00 + 0.00 i$ 
    & &
    & $-0.02 + 0.00 i \phantom{-}$
    \\
    $X_{\pi ^{+} \Xi ^{-}}$ & $0.00 + 0.00 i$ 
    & &
    & $0.00 + 0.00 i$
    \\
    $X_{\pi ^{0} \Xi ^{0}}$ & $0.00 + 0.00 i$ 
    & &
    & $0.00 + 0.00 i$ 
    \\
    $X_{\eta \Xi ^{0}}$ & $0.02 + 0.01 i$ 
    & &
    & $0.01 + 0.02 i$
    \\
    $Z$ & $0.08 + 0.04 i$ 
    & &
    & $0.06 + 0.11 i$ 
    \\
    \hline
    \hline
  \end{tabular}
\end{table}

As we have explained, we have four subtraction constants as the model
parameters.  First we fix them by using the so-called natural
renormalization scheme~\cite{Hyodo:2008xr}, which can exclude explicit
pole contributions from the loop functions.  In the natural
renormalization scheme, we introduce a certain energy $w_{\rm m}$ at
which we achieve the consistency of the low-energy theorem with
respect to the spontaneous breaking of the chiral symmetry.  Namely,
since we take the interaction $V$ as the leading order term of the
chiral perturbation theory, we require that the scattering amplitude
$T$ should coincide with the interaction $V$ at certain ``low'' energy
according to the low-energy theorem.  We represent this energy as
$w_{\rm m}$: $T_{j k} ( w_{\rm m} ) = V_{j k} ( w_{\rm m} )$ with
$G_{j} ( w_{\rm w} ) = 0$ in every channel $j$.  According to the
discussion in Ref.~\cite{Hyodo:2008xr}, we fix this matching energy
scale as the mass of the ``target'' baryon of the scatterings, i.e.,
$w_{\rm m} = M_{\Lambda}$, $M_{\Sigma}$, or $M_{\Xi}$.\footnote{We
  note that the energy $w_{\rm m} = M_{\Lambda}$ is on the left-hand
  cut of the $\pi \Xi$ channel.  Nevertheless, we employ this energy
  as the matching scale, since the $\pi \Xi$ contribution to $\Xi
  (1690)$ is found negligible.}  As a result, we obtain the
subtraction constants in the second, third, and fourth columns of
Table~\ref{tab:neutral}, to which we refer as the parameter
sets~$\Lambda$, $\Sigma$, and $\Xi$, respectively.\footnote{Since we
  assume the isospin symmetry for the subtraction constants, these
  subtraction constants are obtained with isospin symmetric masses for
  hadrons in the natural renormalization scheme.}  With the parameter
set~$\Lambda$, we find two resonance poles as $\Xi ^{\ast}$ states
with $J^{P} = 1/2^{-}$; each pole is in the second (unphysical)
Riemann sheets of the open channels, whose thresholds are lower than
$\text{Re} (w_{\rm pole})$.  One pole is located at $1556.5 - 102.9 i
\mev$, which corresponds to the pole studied in
Ref.~\cite{Ramos:2002xh, GarciaRecio:2003ks, Gamermann:2011mq} as the
$\Xi (1620)$ resonance.  We have found that the energy dependence of
the Weinberg--Tomozawa interaction in the $\pi \Xi$ channel is
essential to the appearance of $\Xi (1620)$.  Actually, by taking into
account only the $\pi \Xi$ channel and switching off couplings to
other channels, we obtain a resonance pole at a similar position in
the $\pi \Xi$ dynamics.  The mechanism is the same as that of the
broad $\Lambda (1405)$ pole in the chiral unitary approach, to which
the energy dependence of the Weinberg--Tomozawa interaction in the
$\pi \Sigma$ channel is essential.  In addition to the $\Xi (1620)$
pole, another pole appears at $1682.6 - 0.8 i \mev$ just below the
$\bar{K} \Sigma$ threshold, whose properties are listed in the second
column of Table~\ref{tab:neutral}.  We expect that the latter pole can
be identified with the $\Xi (1690)$ resonance and originates from the
$\bar{K} \Sigma$ bound state.  On the other hand, with the parameter
sets~$\Sigma$ and $\Xi$, we obtain no poles near the $\bar{K} \Sigma$
threshold as $\Xi (1690)$.  However, we will not take these parameter
sets seriously, since they give larger value of the ratio $R$ with
larger $\chi ^{2}$ value introduced below and can be excluded by the
experimental results.

Let us now concentrate on the $\Xi ^{\ast}$ state near the $\bar{K}
\Sigma$ threshold in the parameter set~$\Lambda$.  The structure of
the $\Xi ^{\ast}$ state can be investigated with the coupling
constants and compositeness.  Actually, from Table~\ref{tab:neutral}
the coupling constants and compositeness indicate a large $\bar{K}
\Sigma$ component in the $\Xi ^{\ast}$ state.  Especially the $\bar{K}
\Sigma$ compositeness, $X_{K^{-} \Sigma ^{+}} + X_{\bar{K}^{0} \Sigma
  ^{0}}$, dominates the sum rule~\eqref{eq:norm} with its small
imaginary part.  This result strongly indicates that the $\Xi ^{\ast}$
state is indeed a $\bar{K} \Sigma$ molecular state, on the basis of
the similarity to the bound state case; the wave function of the $\Xi
^{\ast}$ state can be similar to that of a bound state dominated by
the $\bar{K} \Sigma$ channel.  We also note that, although the
coupling constant approximately satisfies the isospin relation
$g_{K^{-} \Sigma ^{+}} = - \sqrt{2} g_{\bar{K}^{0} \Sigma ^{0}}$, the
$\bar{K} \Sigma$ compositeness largely breaks the corresponding
relation $X_{K^{-} \Sigma ^{+}} = 2 X_{\bar{K}^{0} \Sigma ^{0}}$.
This is because the $\Xi ^{\ast}$ state is located very close to the
$K^{-} \Sigma ^{+}$ threshold.  This fact will induce further effects
of the isospin symmetry breaking on $\Xi (1690)$, such as the
difference of the $K^{-} \Sigma ^{+}$ and $\bar{K}^{0} \Sigma ^{0}$
mass spectra, due to the difference of their thresholds.  The
dominance of the $K^{-} \Sigma ^{+}$ compositeness implies a
coupled-channels extension of the near-threshold scaling in $s$ wave
discussed in Refs.~\cite{Hyodo:2014bda, Hanhart:2014ssa}.

In addition to the $\bar{K} \Sigma$ component, the $\Xi ^{\ast}$ state
has a remarkable property of its small decay width with the small
imaginary part of the pole position $\sim 1 \mev$, which can be seen
also in Refs.~\cite{GarciaRecio:2003ks, Gamermann:2011mq}.  This decay
property can be understood by considering the structure of the
coefficient $C_{j k}$.  Namely, as shown in Table~\ref{tab:coff0}, the
transitions $K^{-} \Sigma ^{+}$, $\bar{K}^{0} \Sigma ^{0}
\leftrightarrow \bar{K}^{0} \Lambda$ are forbidden at the leading
order, so the decay of the $\bar{K} \Sigma$ quasibound state to the
$\bar{K} \Lambda$ channel is highly suppressed.  Actually, the decay
to $\bar{K} \Lambda$ takes place through a multiple scattering of
$\bar{K} \Sigma \to \eta \Xi \to \bar{K} \Lambda$, since the $\bar{K}
\Sigma$-$\eta \Xi$ coupling is the strongest among the coupled
channels~\cite{Ramos:2002xh}.  In addition, the $\bar{K} \Sigma
\leftrightarrow \pi \Xi$ transition is not strong compared to, {\it
  e.g.}, the $\bar{K} N \leftrightarrow \pi \Sigma$ one in the
$\Lambda (1405)$ case; the coefficient $C_{\bar{K} \Sigma \, \pi \Xi
  (I=1/2)}$ in the isospin basis is $- 1/2$~\cite{Ramos:2002xh}, while
that of $\bar{K} N \leftrightarrow \pi \Sigma$ in $I = 0$ is $-
\sqrt{3/2}$~\cite{Oset:1997it}.  As a consequence, the $\Xi ^{\ast}$
state as a $\bar{K} \Sigma$ molecule cannot couple strongly to
$\bar{K} \Lambda$ nor $\pi \Xi$ as the decay channels and hence the
decay width becomes very small.  Moreover, the above argument can also
explain the small ratio of the $\Xi (1690)$ branching fractions
$\Gamma (\pi \Xi) / \Gamma (\bar{K} \Sigma) <
0.09$~\cite{Agashe:2014kda}.  Here we note that higher-order
contributions to the interaction, such as the $s$- and $u$-channel
Born terms, can bring tree-level couplings of $\bar{K} \Sigma
\leftrightarrow \bar{K} \Lambda$ and may give a decay width $\lesssim
10 \mev$.

\begin{figure}[!t]
  \centering
  \Psfig{7.7cm}{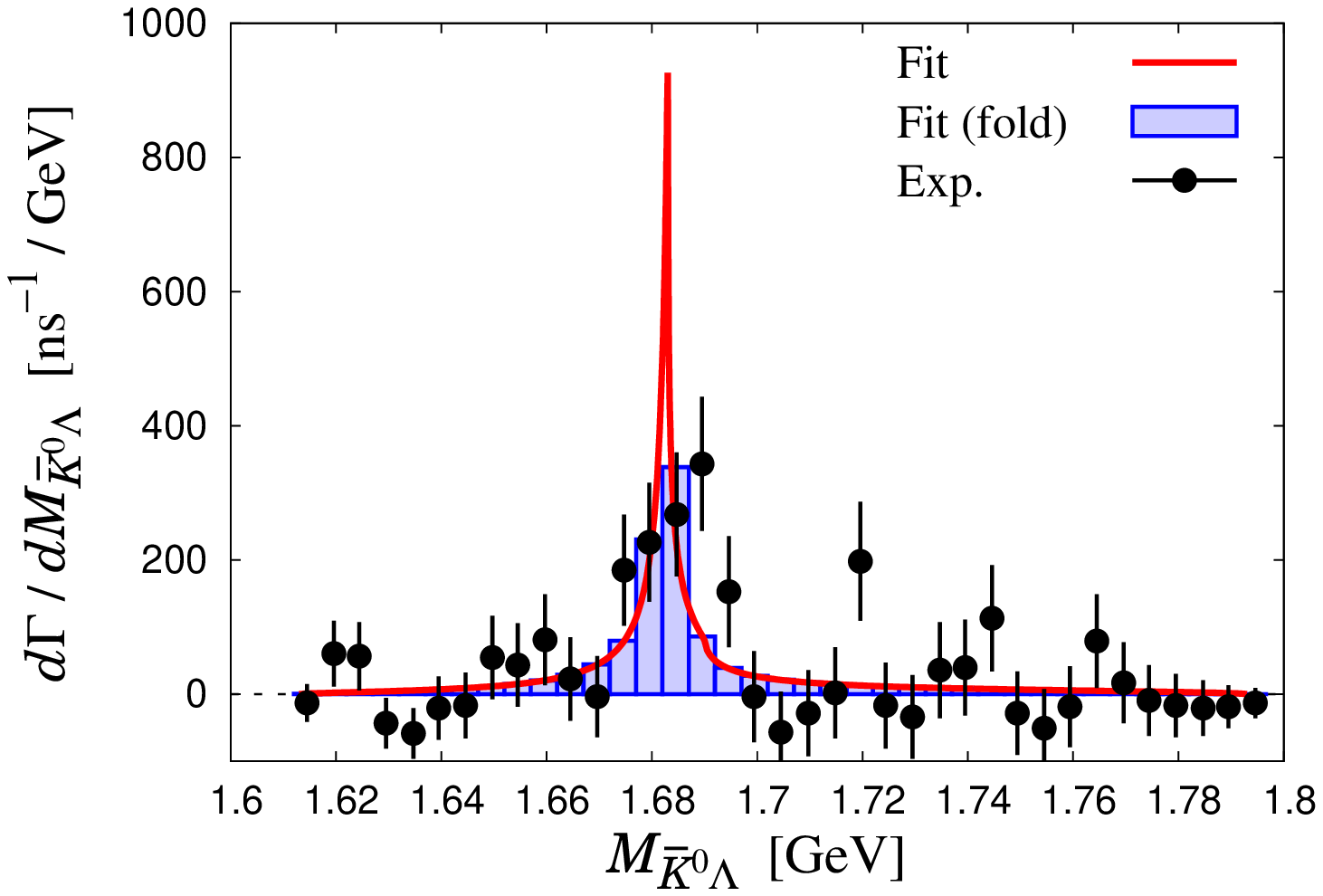}~\Psfig{7.7cm}{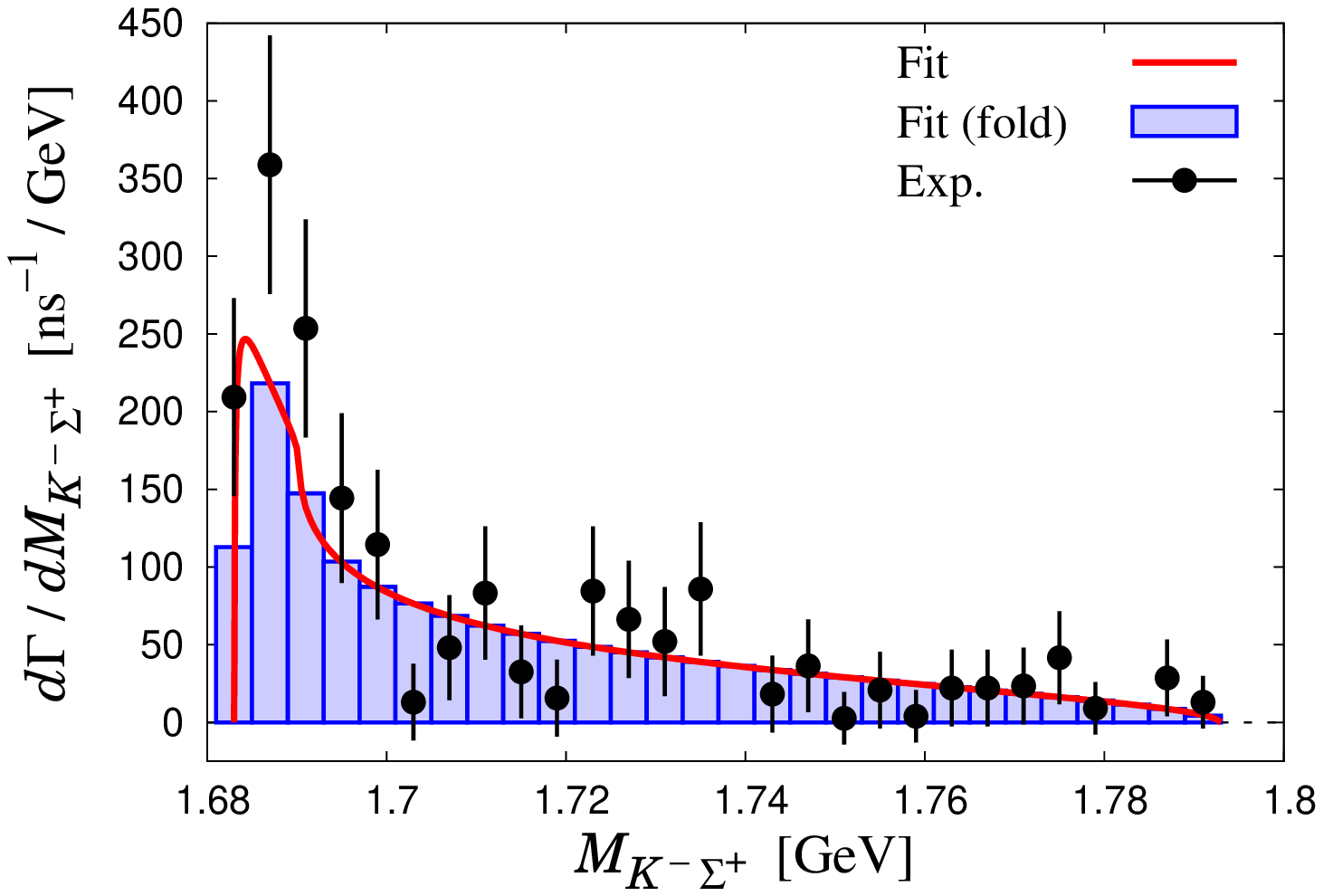}
  \caption{Mass spectra of $\bar{K}^{0} \Lambda$ (left) and $K^{-}
    \Sigma ^{+}$ (right) calculated with Eq.~\eqref{eq:dGdM}.  The red
    solid lines are the theoretical mass spectra with the fitted
    parameters and $C = 0.222 \text{ ns}^{-1}/\text{GeV}$, and the
    blue histograms correspond to the folded results with the size of
    experimental bins, $5 \mev$ and $4 \mev$ for $\bar{K}^{0} \Lambda$
    and $K^{-} \Sigma ^{+}$, respectively.  Experimental data are
    taken from Ref.~\cite{Abe:2001mb}, and their scale is fixed with
    the experimental branching fractions (see text). }
  \label{fig:KY}
\end{figure}

Next, in order to make things more accurate, we fit the $\bar{K}^{0}
\Lambda$ and $K^{-} \Sigma ^{+}$ mass spectra to the experimental data
taken from the decay processes $\Lambda _{c}^{+} \to \Xi (1690)^{0}
K^{+} \to ( \bar{K}^{0} \Lambda ) K^{+}$ and $(K^{-} \Sigma ^{+})
K^{+}$ in Ref.~\cite{Abe:2001mb}.  The scale of two mass spectra is
fixed with the central values of $\mathcal{B} [\Lambda _{c}^{+} \to
\Xi (1690)^{0} K^{+} \to ( \bar{K}^{0} \Lambda ) K^{+}] = ( 1.3 \pm
0.5 ) \times 10^{-3}$ and $\mathcal{B} [\Lambda _{c}^{+} \to \Xi
(1690)^{0} K^{+} \to ( K^{-} \Sigma ^{+} ) K^{+}] = ( 8.1 \pm 3.0 )
\times 10^{-4}$~\cite{Agashe:2014kda}, respectively.  In this study,
according to Ref.~\cite{Flatte:1976xu}, we calculate the two mass
spectra with the correct phase-space factor and a constant prefactor
$C$ as
\begin{equation}
\frac{d \Gamma _{( \bar{K}^{0} \Lambda ) K^{+}}}
{d M_{\bar{K}^{0} \Lambda}} 
= 
C p_{K} p_{\Lambda}^{\ast} 
| T_{\bar{K} \Sigma (I=1/2) \to \bar{K}^{0} \Lambda} |^{2} ,
\quad 
\frac{d \Gamma _{( K^{-} \Sigma ^{+} ) K^{+}}}
{d M_{K^{-} \Sigma ^{+}}} 
= 
C p_{K} p_{\Sigma}^{\ast} 
| T_{\bar{K} \Sigma (I=1/2) \to K^{-} \Sigma ^{+}} |^{2} ,
\label{eq:dGdM}
\end{equation}
where $M_{\bar{K} Y}$ is the invariant mass of the $\bar{K} Y =
\bar{K}^{0} \Lambda$ or $K^{-} \Sigma ^{+}$ system and $p_{K}$
($p_{Y}^{\ast}$) is the momentum of $K^{+}$ (hyperon $Y$) in the
$\Lambda _{c}^{+}$ ($\bar{K} Y$) rest frame.  The constant $C$ is
common to the two modes, since we expect that both the two mass
spectra are obtained with the decay of $\Xi (1690)$ as a $\bar{K}
\Sigma (I=1/2)$ molecular state, and $C$ is determined by the fitting
procedure together with the subtraction constants.  Moreover, the
scattering amplitude $\bar{K} \Sigma (I=1/2) \to j$ [$j = 1$ ($K^{-}
\Sigma ^{+}$), $3$ ($\bar{K}^{0} \Lambda $)] is calculated as
\begin{equation}
T_{\bar{K} \Sigma (I=1/2) \to j}
= - \sqrt{\frac{2}{3}} T_{1 j} 
+ \sqrt{\frac{1}{3}} T_{2 j} , 
\end{equation}
where we have used the $\bar{K} \Sigma ( I = 1/2 )$ state in our
convention
\begin{equation}
| \bar{K} \Sigma ( I = 1/2, \, I_{z} = 1/2) \rangle
= - \sqrt{\frac{2}{3}} | K^{-} \Sigma ^{+} \rangle
+ \sqrt{\frac{1}{3}} | \bar{K}^{0} \Sigma ^{0} \rangle .
\label{eq:KSigma}
\end{equation}
From the best fit to the $\bar{K}^{0} \Lambda$ and $K^{-} \Sigma ^{+}$
mass spectra shown in Fig.~\ref{fig:KY}, we obtain the subtraction
constants in the fifth column of Table~\ref{tab:neutral} (Fit).  With
the parameter set~Fit, a $\Xi ^{\ast}$ state is dynamically generated
as a pole at $1684.3 - 0.5 i \mev$ as in Table~\ref{tab:neutral}.
Since this $\Xi ^{\ast}$ pole qualitatively reproduces the $\Xi
(1690)$ peak in the mass spectra as shown in Fig.~\ref{fig:KY}, we can
identify this $\Xi ^{\ast}$ state with the $\Xi (1690)$ resonance.

An important point for the mass spectra in Fig.~\ref{fig:KY} is that,
although we employ the simplest interaction, i.e., the
Weinberg--Tomozawa interaction, our scattering amplitude qualitatively
reproduces the experimental $\Xi (1690)^{0}$ peaks in both the
$\bar{K}^{0} \Lambda$ and $K^{-} \Sigma ^{+}$ mass spectra.
Especially we emphasize that the $K^{-} \Sigma ^{+}$ mass spectrum
shows a rapid enhancement at its threshold, which is a consequence of
the fact that there is $\Xi (1690)$ near the $K^{-} \Sigma ^{+}$
threshold.  The peak height of the enhancement reflects how strong
$\Xi (1690)$ couples to the $K^{-} \Sigma ^{+}$ channel, or in other
words how much $\Xi (1690)$ contains the $K^{-} \Sigma ^{+}$
component.  In this sense, it is essential to observe both the
$\bar{K} \Sigma$ and $\bar{K}^{0} \Lambda$ mass spectra and to
determine the relative strength between them in experiments.  The
enhancement of the $K^{-} \Sigma ^{+}$ mass spectrum can be evaluated
with the ratio of the two branching fractions as
\begin{equation}
  R \equiv \frac{\mathcal{B} [\Lambda _{c}^{+} \to \Xi (1690)^{0} K^{+} 
    \to ( K^{-} \Sigma ^{+} ) K^{+}]}
  {\mathcal{B} [\Lambda _{c}^{+} \to \Xi (1690)^{0} K^{+} 
    \to ( \bar{K}^{0} \Lambda ) K^{+}]} ,
  \label{eq:R}
\end{equation}
whose experimental value is $[ ( 8.1 \pm 3.0 ) \times 10^{-4} ] / [(
1.3 \pm 0.5 ) \times 10^{-3}] = 0.62 \pm 0.33$~\cite{Agashe:2014kda}.
Theoretically $R$ is obtained by the ratio of integrals of the two
mass spectra, and the result is shown in Table~\ref{tab:neutral}.  The
theoretical values of $R$ overestimate the experimental one, and
especially the parameter sets~$\Sigma$ and $\Xi$ can be excluded.  On
the other hand, the $R$ value of the set~Fit is in $2 \sigma$ errors
of the experimental value, although the statistical error is not small
for the experimental value.  Therefore, an experimental determination
of $R$ can constrain more the $\bar{K} \Sigma$ scattering amplitude
and the structure of the $\Xi (1690)$ resonance.

We note that the $\Xi (1690)$ pole in the parameter set~Fit is located
in the first Riemann sheet of the $K^{-} \Sigma ^{+}$, $\bar{K}^{0}
\Sigma ^{0}$, and $\eta \Xi ^{0}$ channels and in the second Riemann
sheet of the $\bar{K}^{0} \Lambda$, $\pi ^{+} \Xi ^{-}$, and $\pi ^{0}
\Xi ^{0}$ channels.  Therefore, this pole exists above the $K^{-}
\Sigma ^{+}$ threshold ($1683.05 \mev$) but in the first Riemann sheet
of this channel, which is connected smoothly from the pole position of
parameter set~$\Lambda$.  In this meaning, strictly speaking, the peak
seen in the $\bar{K}^{0} \Lambda$ mass spectrum in Fig.~\ref{fig:KY}
is a cusp at the $K^{-} \Sigma ^{+}$ threshold rather than the usual
Breit--Wigner resonance peak.  Other properties of $\Xi (1690)$ in the
parameter set~Fit are very similar to those in the parameter
set~$\Lambda$.  The $\pi \Xi$ subtraction constant in the set~Fit is
larger than ``natural'' value $\sim -2$~\cite{Oller:2000fj}, but we do
not take it seriously since the $\pi \Xi$ channel negligibly
contribute to the $\Xi (1690)$ resonance.  Actually, even changing the
subtraction constant $a_{\pi \Xi} = -0.75$ to $-2$ in the set~Fit, we
obtain a similar $\chi ^{2}$ value.

\section{Discussion}

\begin{figure}[t]
  \centering
  \begin{minipage}{0.54\hsize}
    \Psfig{7.7cm}{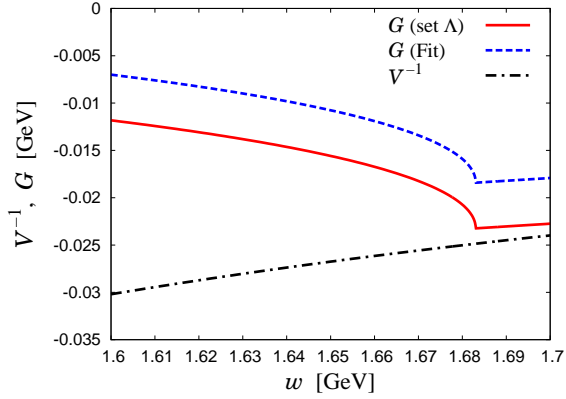} 
  \end{minipage}
  \begin{minipage}{0.45\hsize}
    \caption{Loop function $G$ in the $K^{-} \Sigma ^{+}$ channel and
      inverse of the interaction $V$ in the $\bar{K} \Sigma (I=1/2)$
      channel [see Eq.~\eqref{eq:V_KS}].  The $K^{-} \Sigma ^{+}$
      threshold is located at $1683.05 \mev$.}
    \label{fig:VG}
  \end{minipage}
\end{figure}

Next we investigate the origin of the $\Xi (1690)$ pole in our
scattering amplitude.  First we expect that, from the value of the
$\bar{K} \Sigma$ compositeness, the $\Xi ^{\ast}$ state originates
from a $\bar{K} \Sigma$ bound state.  In order to check this, we
clarify whether the chiral $\bar{K} \Sigma$ interaction in the isospin
$I = 1/2$ is attractive enough to generate a bound state when the
couplings to other channels are switched off.  By using the $\bar{K}
\Sigma$ state in the isospin basis in Eq.~\eqref{eq:KSigma}, the
$\bar{K} \Sigma$ interaction in $I = 1/2$ is expressed as
\begin{equation}
V_{\bar{K} \Sigma (I=1/2)} 
= \frac{2}{3} V_{1 1} - \frac{\sqrt{2}}{3} ( V_{1 2} + V_{2 1} )
+ \frac{1}{3} V_{2 2} .
\label{eq:V_KS}
\end{equation}
In case of a single channel problem, a bound state appears at the
energy which satisfies $V^{-1} = G$ below the threshold.  Therefore,
we compare the behavior of the loop function $G$ in the $K^{-} \Sigma
^{+}$ channel and the inverse of the $\bar{K} \Sigma (I=1/2)$
interaction $[V_{\bar{K} \Sigma (I=1/2)}]^{-1}$, which are plotted in
Fig.~\ref{fig:VG}.  A bound state would appear at the energy where the
lines $G$ and $V^{-1}$ intersect each other in Fig.~\ref{fig:VG}, but
in fact there is no intersection as the inverse of the interaction
$V^{-1}$ is a bit below the loop function $G$.  This means that the
chiral $\bar{K} \Sigma$ interaction is attractive but not enough to
generate a bound state in a single channel problem.  This is in
contrast to the $\bar{K} N (I=0)$ interaction, which can solely
generate a bound state as the origin of the $\Lambda (1405)$
resonance~\cite{Hyodo:2007jq}.  In addition, this fact implies that
the multiple scatterings such as $\bar{K} \Sigma \to \eta \Xi \to
\bar{K} \Sigma$ assist the $\bar{K} \Sigma$ interaction in dynamically
generating a $\bar{K} \Sigma$ quasibound state which is located very
close to the $\bar{K} \Sigma$ threshold.  In the multiple scatterings
the $\eta \Xi$ channel will be the most important, since the
coefficient $C_{j k}$ of the $\bar{K} \Sigma$-$\eta \Xi$ coupling is
the strongest among the coupled channels.  This can be seen also from
the large coupling constant $g_{\eta \Xi}$ in Table~\ref{tab:neutral},
which is comparable to the $\bar{K} \Sigma$ coupling constants.  Here
we note that, when the kaon decay constant is chosen to be $f_{K} =
f_{\pi} \approx 90 \mev$, attraction of the $\bar{K} \Sigma$
interaction will become stronger and the $\bar{K} \Sigma$ interaction
may be able to solely generate a bound state.  As a result, in this
condition the binding energy of the $\bar{K} \Sigma$ system as $\Xi
(1690)$ will be several tens of MeV, which was indeed achieved in
Ref.~\cite{GarciaRecio:2003ks}.  Furthermore, the subtraction constant
$a_{\eta \Xi} $ is negatively large compared to the values in the
natural renormalization scheme.  This would reflect effects from
implicit channels such as $\bar{K}^{\ast} \Sigma$, which were taken
into account in Ref.~\cite{Gamermann:2011mq}.

\def\arraystretch{1.0}

\begin{table}[!t]
  \caption{Properties of the charged $\Xi (1690)$ state with the 
    parameter set ``Fit'' for the neutral $\Xi (1690)$ state.
    The pole position is $w_{\rm pole} = 1693.4 - 10.5 i \mev$.}
  \label{tab:charged}
  \centering
  \begin{tabular}{lcclc}
    \hline
    \hline
    \multicolumn{5}{c}{Fit}
    \\
    \hline
    $g_{\bar{K}^{0} \Sigma ^{-}}$
    & $2.17 + 0.29 i$ 
    & &
    $X_{\bar{K}^{0} \Sigma ^{-}}$ 
    & $0.86 - 0.50 i$
    \\
    $g_{K^{-} \Sigma ^{0}}$
    & $1.36 + 0.07 i$ 
    & &
    $X_{K^{-} \Sigma ^{0}}$ 
    & $-0.27 + 0.31 i \phantom{-}$
    \\
    $g_{K^{-} \Lambda}$ 
    & $0.76 + 0.04 i$ 
    & &
    $X_{K^{-} \Lambda}$ 
    & $-0.02 + 0.04 i \phantom{-}$
    \\
    $g_{\pi ^{-} \Xi ^{0}}$ 
    & $0.18 - 0.09 i$ 
    & &
    $X_{\pi ^{-} \Xi ^{0}}$ 
    & $0.00 + 0.00 i$
    \\
    $g_{\pi ^{0} \Xi ^{-}}$ 
    & $-0.07 - 0.20 i \phantom{-}$
    & &
    $X_{\pi ^{0} \Xi ^{-}}$ 
    & $0.00 + 0.00 i$ 
    \\
    $g_{\eta \Xi ^{-}}$ 
    & $-1.41 - 0.33 i \phantom{-}$
    & &
    $X_{\eta \Xi ^{-}}$ 
    & $0.07 + 0.03 i$
    \\
    & & & 
    $Z$ 
    & $0.36 + 0.12 i$
    \\
    \hline
    \hline
  \end{tabular}
\end{table}

Finally we consider the charged $S = -2$ system with the channels
$\bar{K}^{0} \Sigma ^{-}$, $K^{-} \Sigma ^{0}$, $K^{-} \Lambda$, $\pi
^{-} \Xi ^{0}$, $\pi ^{0} \Xi ^{-}$, and $\eta \Xi ^{-}$.  Here we use
the parameter set~Fit given in Table~\ref{tab:neutral} for the
subtraction constants and solve the Bethe--Salpeter
equation~\eqref{eq:BSeq} to obtain the scattering amplitude.  As a
result, with the set~Fit we find a pole near the $\bar{K} \Sigma$
threshold, which corresponds to the $\Xi (1690)^{-}$ resonance.  The
pole is located in the first Riemann sheet of the $\bar{K}^{0} \Sigma
^{-}$ and $\eta \Xi ^{-}$ channels and in the second Riemann sheet of
the $K^{-} \Lambda$, $K^{-} \Sigma ^{0}$, $\pi ^{-} \Xi ^{0}$, and
$\pi ^{0} \Xi ^{-}$ channels.  The properties of $\Xi (1690)^{-}$ are
listed in Table~\ref{tab:charged}.  The pole position has a larger
imaginary part $\sim 10 \mev$ compared to the neutral case, since it
exists above the $K^{-} \Sigma ^{0}$ threshold in its second Riemann
sheet and hence the decay $\Xi (1690)^{-} \to K^{-} \Sigma ^{0}$ is
allowed.  The coupling constants and compositeness indicate that $\Xi
(1690)^{-}$ has a large $\bar{K} \Sigma$ component.  However, each of
$\bar{K}^{0} \Sigma ^{-}$ and $K^{-} \Sigma ^{0}$ compositeness has a
nonnegligible imaginary part, because the pole exists above the $K^{-}
\Sigma ^{0}$ one in its second Riemann sheet.  Nevertheless, the sum
$X_{\bar{K}^{0} \Sigma ^{-}} + X_{K^{-} \Sigma ^{0}}$ is the largest
contribution to the sum rule~\eqref{eq:norm} with its small imaginary
part, which implies that the charged $\Xi ^{\ast}$ state is also a
$\bar{K} \Sigma$ molecular state.  Moreover, we can expect effects of
the isospin symmetry breaking on the charged $\Xi (1690)$ state in a
similar manner to the neutral case due to the $K^{-} \Sigma
^{0}$-$\bar{K}^{0} \Sigma ^{-}$ threshold difference.

\section{Conclusion}

In this study we have investigated dynamics of $\bar{K} \Sigma$ and
its coupled channels in the chiral unitary approach.  It is a great
advantage to employ the chiral unitary approach that a simple
interaction kernel provided by the leading-order chiral perturbation
theory does not contain free parameters and hence only the subtraction
constants (or cut-offs) of the loop functions are the model
parameters.  The subtraction constants are fixed in the natural
renormalization scheme, which can exclude explicit pole contributions
from the loop functions, or by fitting the $\bar{K}^{0} \Lambda$ and
$K^{-} \Sigma ^{+}$ mass spectra to the experimental data.

As a result, we have found that, although the $\bar{K} \Sigma$
interaction from the leading-order chiral perturbation theory alone is
slightly insufficient to bring a $\bar{K} \Sigma$ bound state,
multiple scatterings in a meson--baryon coupled-channels approach can
dynamically generate a $\bar{K} \Sigma$ quasibound state near the
$\bar{K} \Sigma$ threshold.  The obtained scattering amplitude can
qualitatively reproduce the experimental data of the $\bar{K}^{0}
\Lambda$ and $K^{-} \Sigma ^{+}$ mass spectra and contains a $\Xi
^{\ast}$ pole as a $\bar{K} \Sigma$ molecule, which can be identified
with the $\Xi (1690)$ resonance.  Due to the small or vanishing
couplings of the $\bar{K} \Sigma$ channel to others, we can naturally
explain the decay properties of $\Xi (1690)$.  We have also pointed
out a possibility to observe effects of the isospin symmetry breaking
on $\Xi (1690)$, such as the difference of the $K^{-} \Sigma ^{+}$ and
$\bar{K}^{0} \Sigma ^{0}$ mass spectra, due to the difference of their
thresholds when the $\Xi (1690)$ pole exists very close to the one of
the $\bar{K} \Sigma$ thresholds.

Finally we suggest that further experimental studies on the $\Xi
(1690)$ resonance and related $\bar{K} \Lambda$ and $\bar{K} \Sigma$
mass spectra are most welcome, since these experiments can constrain
more the $\bar{K} \Sigma$ scattering amplitude and the structure of
the $\Xi (1690)$ resonance.  Especially future studies on
multi-strangeness systems at J-PARC, JLab, and other facilities can
shed light on the structure of the $\Xi (1690)$ resonance.  We also
expect that high-statistics analyses on $\Xi (1690)$ by Belle, BaBar,
and LHCb are promising.  On the other hand, for the theoretical
support in analyzing the $\Xi (1690)$ production, structure, and decay
processes, we expect that we can utilize the same or a similar
approach to the $\Lambda (1405)$ case, which has been extensively
studied in the chiral unitary approach as well as in many other
models, effective theories, and lattice QCD simulations.  This is
because both are dynamically generated resonances in the meson--baryon
degrees of freedom and especially originate from the same flavor
$SU(3)$ multiplet in the chiral unitary approach.

\ack

The author greatly acknowledges K.~Imai for suggesting the author to
study this topic and for useful discussions.  The author also thanks
Y.~Kato for useful comments on Belle data and J.~Nieves on the $\Xi
^{\ast}$ states in the chiral unitary approach.  Stimulating
discussions with A.~Hosaka and his careful reading of the manuscript
are gratefully acknowledged.
This work is partly supported by the Grants-in-Aid for Young
Scientists from JSPS (No.~15K17649) and for JSPS Fellows
(No.~15J06538).

\end{document}